\documentclass[letterpaper, 10 pt, conference]{ieeeconf}  
\IEEEoverridecommandlockouts                              
\title{\LARGE \bf
Parameter Estimation in Ill-conditioned Low-inertia Power Systems
}

	\author{Rajasekhar~Anguluri,~\IEEEmembership{Member,~IEEE,}
		Lalitha~Sankar,~\IEEEmembership{Senior~Member,~IEEE,}
		and~Oliver~Kosut,~\IEEEmembership{Member,~IEEE}
		\thanks{This work is funded in part by the NSF under grant OAC-1934766. All the authors are with the School of Electrical, Computer, and Energy
			Engineering, Arizona State University, Tempe, AZ, 85281 USA (e-mail:
			\{rangulur,lalithasankar,okosut\}@asu.edu).}}
			
\usepackage{graphics,color}
\usepackage[outdir=./]{epstopdf}
\usepackage{amsmath}
\usepackage{amssymb}
\usepackage{mathrsfs}
\usepackage{subfigure}
\usepackage[font=small,skip=2pt]{caption}
\usepackage{url}
\usepackage{float}
\usepackage{bbm}
\usepackage{booktabs}
\usepackage{array}
\usepackage[table]{xcolor}
\usepackage{optidef} 
\def\BibTeX{{\rm B\kern-.05em{\sc i\kern-.025em b}\kern-.08em
		T\kern-.1667em\lower.7ex\hbox{E}\kern-.125emX}}

\newtheorem{remark}{Remark}

\newcommand{\map}[3]{#1: #2 \rightarrow #3}
\newcommand{\setdef}[2]{\{#1 \; : \; #2\}}
\newcommand{\subscr}[2]{{#1}_{\textup{#2}}}
\newcommand{\supscr}[2]{{#1}^{\texƒtup{#2}}}
\newcommand{\until}[1]{\{1,\dots,#1\}}
\newcommand{\fromto}[2]{\{#1,\dots,#2\}}
\newcommand{\blkdiag}{\textup{blkdiag}}

\newcommand{\todo}[1]{\par\noindent{\color{blue}\raggedright\textsc{#1}\par\marginpar{\large$\star$}}}
\newcommand{\margin}[1]{\marginpar{\color{blue}\tiny\ttfamily{FP:} #1}}
\newcommand{\flomargin}[1]{\marginpar{\color{blue}\tiny\ttfamily{Flo:} #1}}
\newcommand{\fpmargin}[1]{\marginpar{\color{red}\tiny\ttfamily{Fabio:} #1}}

\newcommand{\Ker}{\operatorname{Ker}}
\newcommand{\Rank}{\operatorname{Rank}}
\newcommand{\Image}{\operatorname{Im}}
\newcommand{\Vtar}{\mathcal{V}^*}
\newcommand{\V}{\mathcal{V}}
\newcommand{\Star}{\mathcal{S}^*}
\newcommand{\Rtar}{\mathcal{R}^*}
\newcommand{\B}{\mathcal{B}}
\newcommand{\real}{\mathbb{R}}
\newcommand{\complex}{\mathbb{C}}
\newcommand{\transpose}{\mathsf{T}} 
\newcommand{\E}{\operatorname{\mathbb{E}}}
\newcommand{\Basis}{\operatorname{Basis}}
\newcommand{\faultset}{\mathcal{K}}
\newcommand{\mc}{\mathcal}
\newcommand{\strucg}{\subscr{\mc G}{i-o}}
\newcommand{\interg}{\subscr{\mc G}{s}}
\newcommand{\indep}{\rotatebox[origin=c]{90}{$\models$}}
\newcommand*{\QEDB}{\hfill\ensuremath{\blacksquare}}
\newcommand*{\QEDW}{\hfill\ensuremath{\square}}
\newcommand{\diag}{\text{diag}}

\newcommand{\imag}{\hat{j}}
\newcommand{\trace}{\text{tr}}
\newcommand{\mcF}{\mathcal{F}}
\newcommand{\what}{\widehat}
\newcommand{\wtilde}{\widetilde}
\newcommand{\minsymbol}{\textasciicircum}
\DeclareMathOperator*{\argmax}{arg\,max}
\DeclareMathOperator*{\argmin}{arg\,min}
\newcommand{\vect}[1]{\mathbbold{#1}}
\newcommand{\vectorones}[1][]{\vect{1}_{#1}}
\newcommand{\vectorzeros}[1][]{\vect{0}_{#1}}
\newcommand{\lammax}{\lambda_\text{max}}
\newcommand{\wtF}{\widetilde{F}}
\newcommand{\wtSig}{\widetilde{\Sigma}}
\newcommand{\tr}{\mathrm{tr}}
\newcommand{\g}{\beta}
\newcommand{\gd}[1]{{\color{blue}[gd: #1]}}
\newcommand{\po}{{(o)}}
\newcommand{\pz}{{(z)}}
\newcommand{\pl}{{(l)}}
\newcommand{\poz}{{(oz)}}
\newcommand{\pzo}{{(zo)}}
\newcommand{\pzeta}{{(\zeta)}}
\newcommand{\mco}{{\mathcal{O}}}
\newcommand{\mcz}{{\mathcal{Z}}}
\newcommand{\pO}{{(\mco\mco)}}
\newcommand{\pZ}{{(\mcz\mcz)}}
\newcommand{\pOZ}{{(\mco\mcz)}}

\newcommand{\mbf}{\mathbf}
\newcommand{\bfx}{\mathbf{x}}
\newcommand{\bfy}{\mathbf{y}}
\newcommand{\bs}{\boldsymbol}

\newcommand\oprocendsymbol{\hbox{$\square$}}
\newcommand\oprocend{\relax\ifmmode\else\unskip\hfill\fi\oprocendsymbol}
\def\eqoprocend{\tag*{$\square$}}

\newcommand{\norm}[1]{\left\lVert#1\right\rVert}

\renewcommand{\theenumi}{(\roman{enumi})}
\renewcommand{\labelenumi}{\theenumi}

\begin{document}

\maketitle
\thispagestyle{empty}
\pagestyle{empty}

\begin{abstract}

This paper examines model parameter estimation in dynamic power systems whose governing electro-mechanical equations are ill-conditioned or singular. This ill-conditioning is because of converter-interfaced power systems generators' zero or small inertia contribution. Consequently, the overall system inertia decreases, resulting in low-inertia power systems. We show that the standard state-space model based on least squares or subspace estimators fails to exist for these models. We overcome this challenge by considering a least-squares estimator directly on the coupled swing-equation model but not on its 
transformed first-order state-space form. We specifically focus on estimating inertia (mechanical and virtual) and damping constants, although our method is general enough for estimating other parameters. 
Our theoretical analysis highlights the role of network topology on the parameter estimates of an individual generator. For generators with greater connectivity, estimation of the associated parameters is more susceptible to variations in other generator states. Furthermore, we numerically show that estimating the parameters by ignoring their ill-conditioning aspects yields highly unreliable results. 
\end{abstract}


\section{INTRODUCTION}

Accurate knowledge of power system model parameters, including inertia and damping, is essential to assess operating states, perform dynamic simulations, and study stability margins. Recently, with increasing penetration of inverter-based (IB) distributed energy resources (DERs) in the bulk power system, the effective system inertia is decreased, making it challenging to stabilize demand-supply mismatch. Further, this increase in IB-DERs significantly increases the number of unknown system parameters to estimate.

Estimating dynamic parameters of synchronous machines and other network devices and loads, is a classical problem 
\cite{kilgore1931calculation, wright1931determination}. Numerous algorithms have been proposed for parameter estimation, both in the presence and absence of closed-loop controllers using local or wide-area ambient measurements, including \cite{burth1999subset, mikhalev2020bayesian, li2021data, zaker2021new, mitra2020online, gorbunov2019estimation, guo2014adaptive, zhao2019power, foroutan2019generator, kosterev2004hydro, ma2008wide, liu2020d, fan2013extended, wang2017estimating}.
Within this body of work, some approaches use \emph{white box models}, wherein the model structure is completely known and deterministic (e.g., model structures given by Newton's laws, mass and energy conservation principles). In power systems, Heffron-Phillips models (fourth order and beyond) have been a mainstay of estimation algorithms. The opposite extreme are \emph{black box models}, or purely data-driven stochastic models in which no prior knowledge is assumed, and an input/output relation is derived from measurements. Examples include modal analysis, 
dynamic equivalents, and Koopman methods. Although black box methods are extremely useful for wide-area monitoring, they have limited utility for planning, contingency, and stability analysis. 

Another line of research focuses on \emph{grey box models} 
\cite{ljung1998system, melgaard1994identification}. These models combine the advantages of the white and black box approaches to exploit prior knowledge of physical relationships or model structure, where possible, and learning unknown parameters from data.
This methodology is particularly relevant in the context of IB-DERs, which inevitably introduce many unknown parameters \cite{guo2020online, wang2017pmu, liu2020line, nouti2021heterogeneous}. However, there are many shortcomings in the existing literature; most papers: (i) focus on net inertia\footnote{Estimated as a weighted average of single area inertia estimates.} rather than the inertia of each device or each area as they connect to each other;
(ii) focus on estimation in the presence of transient disturbance with little work on ambient disturbances; and (iii) do not consider the effect of frequency and voltage dependent loads, leading to large (inertia) estimation errors (up to 40\% \cite{zografos2016estimation}); see \cite{heylen2021challenges, tan2022power}, for a recent account on various parameter estimation in low-inertia systems. 

We put forth a simple strategy for overcoming the above limitations using a simple constrained least squares estimator to estimate parameters using ambient measurements. Least squares type estimators are already used for estimating inertia in power systems; however, these estimators assume that the inertia is strictly greater than zero. This assumption implies that the electro-mechanical dynamics are well defined. However, this assumption does not hold for converter-based generators. For e.g., droop-control based generators provide zero inertia \cite{nouti2021heterogeneous}. Consequently, the electro-mechanical dynamics are not well-defined or ill-conditioned, thereby giving rise to a descriptor system (see Section III for details). We develop a framework for parameter estimation for these systems, with special attention to inertia and damping. Beyond the motivating example of parameter estimation in power systems, our results apply more broadly to other engineering systems modeled using second-order differential equations, such as structural mechanical and acoustic systems and fluid mechanics. We summarize our contributions below: 
\begin{enumerate}
		\item For low-inertia power systems consisting of synchronous and converter-interfaced generators, we study a constrained least-squares estimation problem that allow us to tackle systems with exactly zero-inertia. 
		\item We highlight the role of network connectivity on the estimation performance. Specifically, using the closed-form formulas of the estimators, we show that for generators with greater connectivity, estimation of the associated parameters is more susceptible to variations in other generator states.
		\item Our simulation results on the IEEE 39 bus system show that estimating the parameters by ignoring their ill-conditioning aspects yields highly unreliable results
		\end{enumerate}

\section{Dynamics of Low-inertia Power Systems}
We introduce the frequency dynamics for a low-inertia power system, comprised of multiple synchronous generators (SGs) and converter-interfaced distributed energy resources (DERs). We later use these models for parameter estimation subject to suitable physical constraints.  

We model a power network of $N$ buses with an undirected graph $\mathcal{G}:=(\mathcal{V}, \mathcal{E})$, where nodes $\mathcal{V}=\{1,\ldots,N\}$ and edges $\mathcal{E}\subseteq \mathcal{V}\times \mathcal{V}$ denote buses and transmission lines, respectively. For a $i$-th node in $\mathcal{V}$ we associate a generator (synchronous or converter-faced) whose frequency response around a steady state is governed by the swing equation \cite{nouti2021heterogeneous, lokhov2018online, lokhov2018uncovering}:
\begin{align}\label{eq: swing eq0}
    \frac{2H_{i}}{\omega_0}\Delta\dot{\omega}_i(t)\!=\!\left[\Delta P_{m,i}(t)\!-\!\Delta P_{e,i}(t)\right]-FC_i(t)+\tilde{\epsilon}_i(t), 
\end{align}
where $\omega_0=120\pi$ is the rated angular frequency, $\Delta\omega_i(t)=\omega_i(t)-\omega_0$, and $2H_i/w_0$ is the inertia constant. $\Delta P_{m,i}(t)$ is the deviation from the steady mechanical power injection. $\Delta P_{e,i}(t)$ is the deviation from the electrical power output, and $\tilde{\epsilon}_i(t)$, a zero-mean Gaussian process with a known variance, models the ambient fluctuation in loads as well as process noise. Further, $\Delta P_{e,i}(t)$ equals the sum of deviations of the power flows on the lines connected to node $i$ \cite{lokhov2018online, lokhov2018uncovering}: 
\begin{align}\label{eq: power balance}
    \Delta P_{e,i}(t)=\sum_{ij \in \mathcal{E}}\Delta P_{ij}(t)=\beta_{ij}(\Delta\delta_i(t)-\Delta\delta_j(t)), 
\end{align}
where $\beta_{ij}\!=\!|V_i||V_j|b_{ij}$ with $b_{ij}\!>\!0$ denoting the susceptance and $|V_i|$, $|V_j|$ are the rated voltage magnitudes. The angular deviation $\Delta\delta_i(t)$ is obtained by integrating $\Delta\omega_i(t)$. 

The frequency controller output $FC_i(t)$ enforces the system frequency stability due to a large imbalance between the mechanical and electrical power. In SGs, primary frequency controllers (PFCs) provide the frequency support. On the other hand, in grid-forming converters, the behavior of PFC is emulated by fast frequency regulators. We assume that this goal is achieved by a proportional feedback control that adjusts the power generation set-point based on the frequency deviation: $FC_i(t)=K_i\Delta\omega_i(t)/w_0$ \cite{nouti2021heterogeneous} (see Remark \ref{ref: control}). 

We assume that some of the loads depend on the system frequency. Similar to frequency controllers, these loads provide a damping stabilizing effect on the frequency. We model these loads as $\Delta P_{i,\text{Load}}(t)=D_{i,\text{load}}\Delta\omega_i(t)/w_0$, where $D_{i,\text{load}}$ is the damping coefficient. By slight abuse of notation, we denote the total frequency support by $FC_{i}(t)=(K_i+D_{i,\text{load}})\Delta\omega_i(t)/w_0$ and let $D_i=K_i+D_{i,\text{load}}$.

We drop $\Delta$ notation in the state variables. From \eqref{eq: swing eq0} and \eqref{eq: power balance}, and our discussions on the frequency controller, we can express the dynamics for all generators compactly as
\begin{align}\label{eq: MIMO system}
\begingroup 
\setlength\arraycolsep{4pt}
\underbrace{\begin{bmatrix}
I & 0\\
0 & M
\end{bmatrix}}_{\triangleq E}
\endgroup 
\begin{bmatrix}
	\dot{\boldsymbol{\delta}}(t)\\
	\dot{\boldsymbol{\omega}}(t)
	\end{bmatrix}\!=\!
	\begingroup 
\setlength\arraycolsep{2pt}
	\underbrace{\begin{bmatrix}
	0 & I \\
	-H_\beta & -D
\end{bmatrix}}_{\triangleq A}
\endgroup 
\begin{bmatrix}
{\boldsymbol{\delta}}(t)\\
{\boldsymbol{\omega}}(t)
\end{bmatrix}\!+\!\begin{bmatrix}
\mbf{0}\\
\boldsymbol{\epsilon}(t)
\end{bmatrix}, 
\end{align}
where ${\boldsymbol{\delta}}=[\delta_1,\ldots,\delta_N]^\transpose \in \mathbb{R}^N$ and $ {\boldsymbol{\omega}},  \dot{\boldsymbol{\delta}}, \dot{\boldsymbol{\omega}}, \boldsymbol{\epsilon}$ and $\mbf{0}$ are defined similarly. The $i$-th component of the process noise $\boldsymbol{\epsilon}(t)$ is given by $\epsilon_i(t)=\tilde{\epsilon}_i(t)+P_{m,i}(t)$. The matrices $I$ and $0$ are $N\times N$ identity and all-zeros matrices. The  Laplacian matrix $H_\beta$ is defined as $[H_\beta]_{ij}=-\beta_{ij}$ for $(i,j)\in \mathcal{E}$ and $[H_g]_{ij}=0$ otherwise; and $[H_g]_{ii}=\sum_{(i,j)\in \mathcal{E}}\beta_{ij}$. Finally, $M=\text{diag}(M_{11},\ldots,M_{ii})$ and $D=\text{diag}(D_{11},\ldots,D_{NN})$ are diagonal inertia and damping matrices, where $D_{ii}=D_i/w_0$ and $M_{ii}=2H_i/w_0$.

From \eqref{eq: MIMO system} note that the Laplacian $H_g$ is determined by the line susceptances, and hence, it is independent of the type of the generator (synchronous or converter-based). Thus, each generator is characterized by $M_i$ and $D_i$. However, we show that the estimates $\hat{M}_i$ and $\hat{D}_i$ are influenced by $H_g$. The effect of $H_g$ is ignored in prior works, which focus on either estimating each machine's inertia or the aggregated inertia. 

The (classical) model in \eqref{eq: MIMO system} is a starting point for many downstream tasks, including control design, storage placement, oscillation localization, and stability analysis. In these applications, the model in \eqref{eq: MIMO system} is simplified by left multiplying $E^{-1}$ on both sides of \eqref{eq: MIMO system}. Unfortunately, in low-inertia power systems, this kind of simplification is not possible because the inertia constant $M_i$ could be small for VSMs and exactly zero for droop-control based generators \cite{?}. Consequently, $E$ in \eqref{eq: MIMO system}
	is not invertible. In this case, this as a \emph{linear descriptor} or \emph{differential-algebraic system}. The latter term derives from the fact that some of the equations represented by \eqref{eq: MIMO system} are purely algebraic (and not differential) in that the left-hand side is zero. These systems appear in the field of robotics, economics, and circuits; in power systems, they also arise when generator dynamics and algebraic power-flow algebraic equations explicitly are considered together. In our case, a descriptor system arises due to the  ill-conditioning of parameters caused by the low inertia of IB-DERs. In the following section, we discuss why parameter estimation is difficult in these systems and then describe a new strategy to obtain reliable parameter estimates.
\begin{remark}\label{ref: control} In general, the frequency controller might not be a simple proportional control and can be of higher order; for example, in SG, turbine dynamics contribute to $FC_i(t)$. However, for ease of analysis, we neglect these dynamics. This approximation is valid for converter-interfaced generators because the controller time constants are small; however, this approximation might not be accurate for SGs. \QEDW
\end{remark}

\section{Structure Preserving Estimation Problem}

For the continuous-time model in \eqref{eq: MIMO system}, we first obtain a discrete-time model using Euler's method. We then formulate a constrained least squares optimization problem for estimating the parameters using this discrete-time model.

We assume that we can estimate the generator states $ {\boldsymbol{\delta}}$ and ${\boldsymbol{\omega}}$ using PMU measurements \cite{?}. Let $k=0,1,\ldots$ and define $\mbf{z}[k]\triangleq \mbf{z}(kT_s)$, where $T_s$ is the discretization step (hereafter, the sampling period) and $\mbf{z}[k]=[ \boldsymbol{\delta}[k]^\transpose \, \boldsymbol{\omega}[k]^\transpose]^\transpose$. The relationship among $T_s$, resolution of the PMU measurements, and the time-scale of the estimation horizon is explained in great detail in \cite{milano2017rotor}. Using the Euler-Mayurama discretization method, we get the discrete-time dynamics  \cite{sarkka2019applied, lokhov2018online}:
    \begin{align}\label{eq: DT-state-space}
E(\mbf{z}[k+1]-\mbf{z}[k])=T_s A\mbf{z}[k]+\begin{bmatrix}
    0\\
    \mbf{r}[k]
    \end{bmatrix}, 
    \end{align}
where $\mbf{r}[k]$ is the discretized process noise (cf. $\boldsymbol{\epsilon}(t)$ in \eqref{eq: MIMO system}), and 
$\mbf{r}[k]\sim \mathcal{N}(0,\Sigma_{\epsilon})$, where  $\Sigma_{\epsilon}=T_s\text{diag}(\sigma^2_1,\ldots,\sigma^2_N)$. The diagonal structure of $\Sigma_{\epsilon}$ is because the ambient fluctuations are spatially uncorrelated across different buses.  

The standard practice in the literature \cite{tamrakar2020inertia, lokhov2018online, lokhov2018uncovering, nouti2021heterogeneous, guo2020online} is to re-write \eqref{eq: DT-state-space} as 
    \begin{align}\label{eq: DT-state-space1}
\mbf{z}[k+1]=(I+T_s E^{-1}A)\mbf{z}[k]+E^{-1}\begin{bmatrix}
    0\\
    \mbf{r}[k]
    \end{bmatrix}, 
    \end{align}
and estimate $A_d\triangleq (I+T_sE^{-1}A)$ using $\mbf{z}[0],\ldots,\mbf{z}[\mathcal{T}-1]$. This na\"ive estimate has many drawbacks: (i) $A_d$ might not be well-defined if $E$ is not invertible, which is the case for droop-control based generators, as discussed earlier; (ii) $E^{-1}$ adversely affects the noise vector by distorting its spatially uncorrelated property; and (iii) decomposing the estimate of $A_d$ to uniquely estimate $M$ and $D$ is impossible in general. 

We overcome the limitations of the na\"ive estimator by considering the following constrained least-squares optimization that does not require $E$ to be invertible: 
\begin{align}\label{eq: inertia_estimation}
\{\hat{M}, \hat{D}\}&=\argmin_{M, D \,\in\, \mathcal{D}}
			\sum_{k=0}^{\mathcal{T}-1}\norm{        E(\mbf{z}[k+1]\!-\!\mbf{z}[k])\!-\!T_sA\,\mbf{z}[k]}_2^2\nonumber\\ &\textrm{ s.t. } 
 			 0\leq D_{ii} \leq D_\text{max}, \text{ for all }i, \\
 			 & \quad \,\,M_i=0, \text{ for } i \in \mathcal{V}_{DC}, \nonumber 
\end{align}
where $\mathcal{D}$ is the set of non-negative diagonal matrices; $D_\text{max}$ is a known term that imposes practical limits on $D$; and $\mathcal{V}_{DC}$ are the nodes corresponding to the droop-control generators. The equality constraint in \eqref{eq: inertia_estimation} ensures that the estimate $\hat{M}_i$, $\text{for } i \in \mathcal{V}_{DC}$, is zero. From \eqref{eq: DT-state-space1}, we note that the expression inside the norm term in \eqref{eq: inertia_estimation} is the process noise $[\mbf{0}^\transpose\, \mbf{r}[k]^\transpose]^\transpose$. Thus, the proposed estimator attempts to find parameters that best explain the variations of the ambient fluctuations over the time horizon $k=0,\ldots, \mathcal{T}-1$. 

We rewrite \eqref{eq: inertia_estimation} in the standard least squares form. Define the vectors $\mbf{m}=[M_{11},\ldots,M_{NN}]^\transpose$, $\mbf{d}=[D_{11},\ldots,D_{NN}]^\transpose$. Let  $\tilde{\boldsymbol{\omega}}[k]=\boldsymbol{\omega}[k+1]-\boldsymbol{\omega}[k]$ and $\boldsymbol{\delta}_{0:\mathcal{T}-1}=[\boldsymbol{\delta}[0],\ldots,\boldsymbol{\delta}[\mathcal{T}-1]]^\transpose$. Let $\text{Diag}(\tilde{\boldsymbol{\omega}}[k])$ be the diagonal matrix with the entries of $\tilde{\boldsymbol{\omega}}[k]$ on the main diagonal, and define the data matrix: 
\begin{align}\label{eq: W structure}
W_{0:\mathcal{T}-1}=\begin{bmatrix}
\text{Diag}(\tilde{\boldsymbol{\omega}}[0]) & T_s\text{Diag}{\boldsymbol{\omega}}[0]\\
\text{Diag}(\tilde{\boldsymbol{\omega}}[1]) & T_s\text{Diag}{\boldsymbol{\omega}}[1]\\
\vdots & \vdots \\
\text{Diag}(\tilde{\boldsymbol{\omega}}[\mathcal{T}-1]) & T_s\text{Diag}{\boldsymbol{\omega}}[\mathcal{T}-1]
\end{bmatrix}. 
\end{align}
Then the optimization in \eqref{eq: inertia_estimation} can be compactly expressed as 
	\begin{equation}\label{eq: inertia_estimation2}
	\hspace{-1.0mm}	\begin{aligned}
\{\hat{\mbf{m}}, \hat{\mbf{d}}\}	&=\argmin_{\mbf{m}, \mbf{d} \,\in\, \mathbb{R}^N}
			\norm{W_{0:\mathcal{T}-1}\begin{bmatrix}
			\mbf{m}\\
			\mbf{d}
			\end{bmatrix}+T_s(I\otimes H_\beta)\boldsymbol{\delta}_{0:\mathcal{T}-1}}_2^2\\ &\textrm{s.t. } 
			\mbf{0} \leq \mbf{d} \leq D_\text{max}\mbf{1}, \text{ and } \Gamma \begin{bmatrix}
			\mbf{m}\\
			\mbf{d}
			\end{bmatrix}=\mbf{0}, 
		\end{aligned}
	\end{equation}
where $\mbf{1}$ is the all-ones vector; $I$ is an $\mathcal{T}\times \mathcal{T}$ identity matrix; and $\otimes$ is the matrix Kronecker product. The $n\times 2N$ selection matrix $\Gamma$ (with $n$ denoting the size of $\mathcal{V}_{DC}$) selects the entries of $\mbf{m}$ associated with the droop-control generators. 

Optimization problems similar to \eqref{eq: inertia_estimation2} are recently studied in the literature of inertia and damping estimation \cite{?}. These studies, however, ignore the zero-inertia constraints and $H_\beta$ term; hence, they require damping constraints to make the estimation problem mathematically well-posed. In contrast, the problem in \eqref{eq: inertia_estimation2} is well-posed even when we ignore the damping constraints, thereby making it useful for the cases where $D_{\max}$ is unknown. Using the special case below, we study the role of topology, encoded in the susceptance matrix $H_b$, on the parametric estimates of the $i$-th generator.

\emph{Special case (unconstrained optimization)}:  Suppose that $W_{0:\mathcal{T}-1}$ has full column rank.\footnote{For an appropriate choice of $N$, in general, the full column rank assumption holds because of the presence of additive noise in the measurements.} Let us ignore the constraints in \eqref{eq: inertia_estimation2}. Then, the problem in \eqref{eq: inertia_estimation2} reduces to the unconstrained least squares problem, which admits the following solution: 
\begin{align}\label{eq: LS-estimate}
\begin{bmatrix}
\hat{\mbf{m}}\\
\hat{\mbf{d}}
\end{bmatrix}&=-T_sW_{0:\mathcal{T}-1}^+(I\otimes H_\beta)\boldsymbol{\delta}_{0:\mathcal{T}-1}, 
\end{align}
where $W_{0:\mathcal{T}-1}^+$ is the pseudo-inverse of $W_{0:\mathcal{T}-1}$, and is given by $W_{0:\mathcal{T}-1}^+=(W_{0:\mathcal{T}-1}^\transpose W_{0:\mathcal{T}-1})^{-1}W_{0:\mathcal{T}-1}^\transpose$.
By exploiting the diagonal structure of the blocks in $W_{0:\mathcal{T}-1}$ in \eqref{eq: W structure}, we can express 
estimates at the $i$-th generator node as 
\begin{align}
\begin{split}\label{eq: simple inertia estimate}
    \hat{m}_i&=-\sum_{j=1}^N [H_\beta]_{i,j}\left(\sum_{k=0}^{\mathcal{T}-1}\left[\frac{c_{i,2}}{c_{i,3}}\tilde{\omega}_i[k]-\frac{c_{i,1}}{c_{i,3}}{\omega}_i[k]\right]\delta_j[k]\right)\\
    \hat{d}_i&=-\sum_{j=1}^N [H_\beta]_{i,j}\left(\sum_{k=0}^{\mathcal{T}-1}\left[\frac{c_{i,0}}{c_{i,3}}{\omega}_i[k]-\frac{c_{i,1}}{c_{i,3}}\tilde{\omega}_i[k]\right]\delta_j[k]\right), 
\end{split}
\end{align} 
where the constants $c_{i,3}=c_{i,0}c_{i,2}-c^2_{i,1}$; $c_{i,0}=\sum_{k=0}^{\mathcal{T}-1}\tilde{\omega}^2_i[k]$; $c_{i,1}=\sum_{k=0}^{\mathcal{T}-1}\tilde{\omega}_i{\omega}_i[k]$; and 
$c_{i,2}=\sum_{k=0}^{\mathcal{T}-1}{\omega}^2_i[k]$. In the above expressions, we set $T_s=1$ for simplicity. The constants $c_{i,0}, c_{i,1}$, and $c_{i,2}$ depend on the $i$-th generator's frequencies. They determine the contribution of the frequency and its rate of change $\tilde{\omega}_i[k]=\omega[k+1]-\omega[k]$ on the $i$-th estimate. 

The $i$-th inertia (or damping) estimate in \eqref{eq: simple inertia estimate} is a weighted average of the suceptance values of the lines connected to the $i$-th node. These weights depend both on the $i$-th node's frequencies and the angles of all generators. Thus, for generators with greater connectivity, estimation of the associated parameters is more susceptible to variations in other generator states. Consequently, these parameters cannot be estimated using local measurements. But it makes sense to estimate the inertia of a largely isolated microgrid as it has a few or no connections with other parts of the network. Finally, we can only estimate the parameters of a generator in a large system when both the local frequency and the power measurements are available. Recall from \eqref{eq: power balance} that the power deviations encode the topological information.

We comment on the statistical properties of the estimate in \eqref{eq: LS-estimate}. Because $\mbf{w}(t)$ and $\boldsymbol{\delta}(t)$ in \eqref{eq: MIMO system} are correlated random processes, $W_{0:\mathcal{T}-1}$ and $\boldsymbol{\delta}_{0:\mathcal{T}-1}$ are random and correlated. Thus, characterizing the distribution of the estimate in \eqref{eq: LS-estimate} is hard. A workaround is to interpret the optimization in \eqref{eq: inertia_estimation2} as a means for obtaining the parameters of the linear model: 
\begin{align}
   -T_s(I\otimes H_\beta)\boldsymbol{\delta}_{0:\mathcal{T}-1}= W_{0:\mathcal{T}-1}\begin{bmatrix}
			\mbf{m}\\
			\mbf{d}
			\end{bmatrix}+\boldsymbol{\zeta}, 
\end{align}
where $\boldsymbol{\zeta}$ and the filtered process noise accumulated over time have same distributions; however, $W_{0:\mathcal{T}-1}$ and $\boldsymbol{\delta}_{0:\mathcal{T}-1}$ are responses due to the initial state. Hence, they are deterministic terms. With these assumptions, it follows that 
\begin{align}\label{eq: statistics of the estimate}
    \begin{bmatrix}
    \hat{\mbf{m}}\\
    \hat{\mbf{d}}
    \end{bmatrix}\!\sim\!\mathcal{N}\left(\begin{bmatrix}
    \mbf{m}^*\\
    \mbf{d}^*
    \end{bmatrix}, {T^2_s}W_{0:\mathcal{T}-1}^+\Sigma_{\boldsymbol{\zeta}} (W_{0:\mathcal{T}-1}^+)^\transpose\right), 
\end{align}
where $(\mbf{m}^*, \mbf{d}^*)$ is the unknown truth, and $\Sigma_{\boldsymbol{\zeta}}$ is the covariance matrix of $\boldsymbol{\zeta}$. The characterization in \eqref{eq: statistics of the estimate} holds even for the non-Gaussian process noise, thanks to the asymptotic (in $\mathcal{T}$) normality of the least squares estimator (see \cite{ljung1998system}). If $\Sigma_{\boldsymbol{\zeta}}$ is diagonal, $W_{0:\mathcal{T}-1}^+\Sigma_{\boldsymbol{\zeta}} (W_{0:\mathcal{T}-1}^+)^\transpose$ is a $2\times 2$ block matrix with diagonal blocks. The off-diagonal blocks capture correlations between the inertia and damping estimate at a given node. Thus, the variance of the estimates are not influenced by the variations in other generator states. Unfortunately, $\Sigma_{\boldsymbol{\zeta}}$ cannot be diagonal because the process noise gets filtered through the dynamics in \eqref{eq: DT-state-space}; and hence, $\Sigma_{\boldsymbol{\zeta}}$ is dense, and so is the covariance matrix in \eqref{eq: statistics of the estimate}. Hence, the variance of each estimate depends both on the network and the variations in other generator states. 


The discrete-time variance $\sigma_i^2T_s$ of $\mbf{r}[k]$ (see \eqref{eq: DT-state-space}) due to the process noise and loads is hidden in the covariance matrix $\text{Cov}(\boldsymbol{\zeta})$. By assuming $\sigma_i^2=\sigma^2$, for all $i\in N$, we can write $\text{Cov}(\boldsymbol{\zeta})=\sigma_iT_sQ$, where the matrix $Q$ solely depends on the system dynamic matrices and the topology. Thus, the covariance term in \eqref{eq: statistics of the estimate} is effectively scaled by $T_s^3\sigma^2$. This result highlights the trade-off between the sampling time and the variance of load fluctuations, allowing us to down- or up-sample measurements to improve the estimates' quality. For instance, for highly fluctuating loads (higher values of $\sigma^2$), we can down-sample the measurements for computational speedup with minimal loss in the estimation performance. 

We close this section by pointing out the importance of the structure preserving estimation problem in \eqref{eq: inertia_estimation} in the case where we have access to a few generator states but not all of them. Here, we cannot rewrite \eqref{eq: inertia_estimation} as the constrained form in \eqref{eq: inertia_estimation2}. Nonetheless, we can use expectation-maximization (EM) type algorithms, which at high level solves the optimization in \eqref{eq: inertia_estimation2}, but the data matrix $W_{0,\mathcal{T}-1}$  should be replaced with Kalman estimates. We leave this study for the future.

\section{Simulation Results}
We illustrate the performance of the structure-persevering inertia and damping constants estimator in \eqref{eq: inertia_estimation2} on the IEEE 39-bus, 10-generator benchmark system. See Ref.~\cite{lokhov2018online} for a single line diagram of the topology and the location of the generator buses. The inertia constants of the generators are summarized in Table \ref{tab:sync-constants} and all damping constants are set to $d_i^*=0.0531$ p.u. We use Kron-reduction technique \cite{lokhov2018online} to obtain the matrix $H_\beta$ in \eqref{eq: MIMO system}. We obtain the initial values of $(\boldsymbol{\delta(t)}, \boldsymbol{\omega(t)})$ and the line susceptance values from \cite{hiskens2013ieee}. We set the discretization time-step $T_s=1/60$ sec. We use these parameter values to generate the frequency measurements using 
the discrete-time model in \eqref{eq: DT-state-space}. 

\begin{table}[H]
\large 
    \caption{IEEE 39-bus synchronous generator inertia (p.u.)}
    \vspace{2.0mm}
    \centering
    \begin{tabular}{ |c|c|c|c|c| } 
 \hline
 $m^*_1$ & $m^*_2$ & $m^*_3$ & $m^*_4$ & $m_5^*$ \\ 
 \hline 
 $0.2228$ & $0.1607$ & $0.1873$ & $0.1517$ & $0.1379$\\
 \hline
  $m^*_6$ & $m^*_7$ & $m^*_8$ & $m^*_9$ & $m_{10}^*$ \\ 
 \hline 
 $0.1846$ & $0.1401$ & $0.18289$ & $0.1830$ & $2.6526$\\
 \hline 
    \end{tabular}
    \label{tab:sync-constants}
\end{table}

\subsection{Case study 1: estimation performance and validation} First, we explore the case where there are no converter-based generators, and the inertia constants of all synchronous generators are not close to zero; see Table \ref{tab:sync-constants}. We also ignore the damping constraints for simplicity. Thus, we consider the unconstrained optimization problem. 

\begin{figure}
		\centering
		\includegraphics[width=1.0\linewidth]{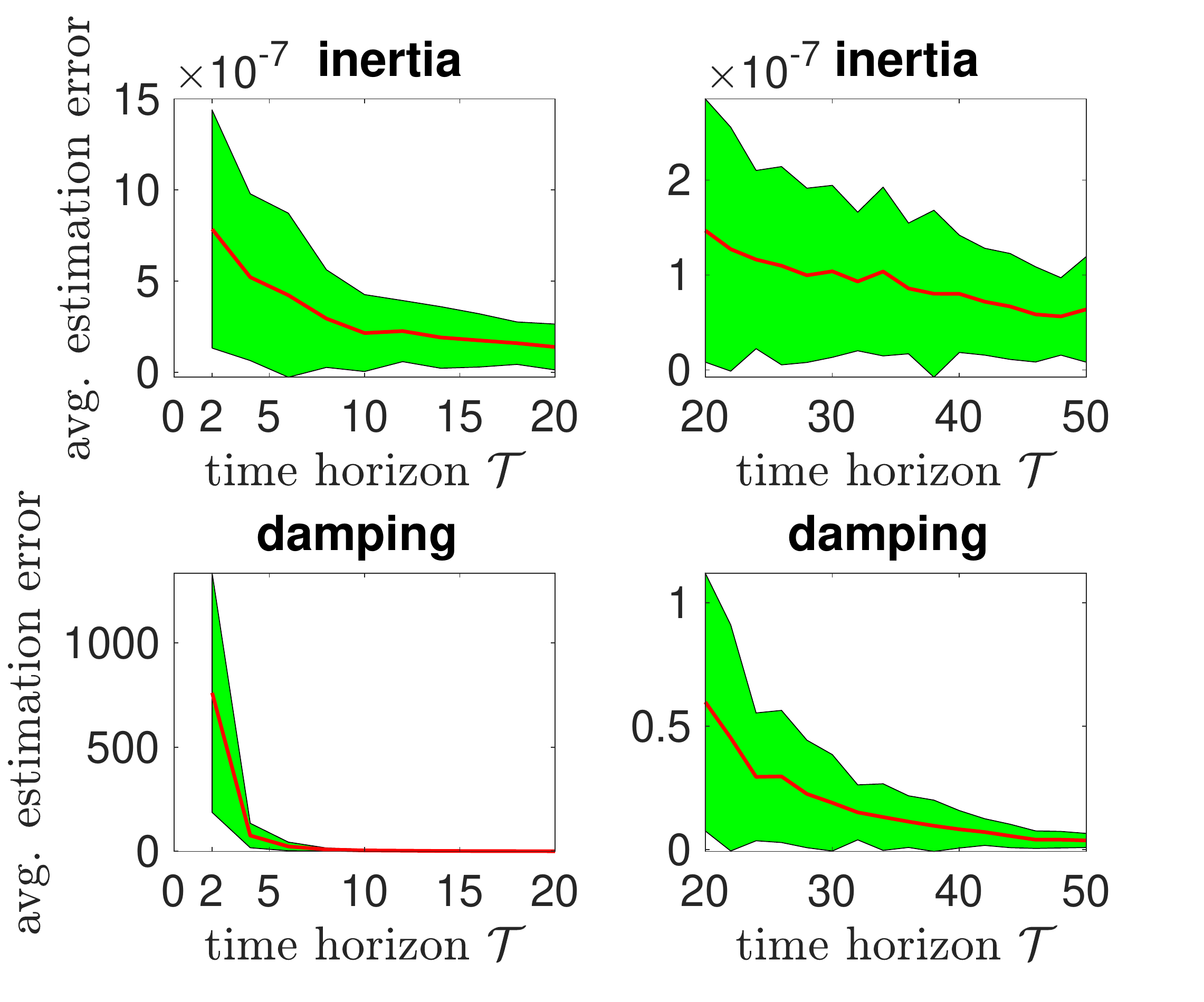} 
		\caption{\small Estimation error as a function of estimation time horizon. The shaded region denotes the standard deviation (averaged over 100 trails). From both the top and bottom panels, we clearly see that the average error in \eqref{eq: error metrics} decreases by increasing $\mathcal{T}$. However, compared to inertia, we need more measurements to estimate damping accurately.} \label{fig: est_error}
	\end{figure}

Our first simulations focus on the inertia and damping estimation error behavior as a function of the estimation time $\mathcal{T}$. We define the following error metrics: 
\begin{align}\label{eq: error metrics}
    E_\text{int}=\frac{1}{10}\sum_{i=1}^{10}(\hat{m}_i-\hat{m}_i^*)^2\,\,; D_\text{int}=\frac{1}{10}\sum_{i=1}^{10}(\hat{d}_i-\hat{d}_i^*)^2. 
\end{align}
These metrics capture estimation error (squared) of a random generator node. We set the process noise standard deviation $\sigma=0.01$ p.u. \cite{lokhov2018online}. Fig.~\ref{fig: est_error} illustrates Montecarlo estimate of the mean and the standard deviation (no. of. trails = 100) of the error metrics. We note that more measurements are required to estimate damping accurately than inertia. This is because the inertia estimate in \eqref{eq: simple inertia estimate} depends more on the difference of the frequencies at $k$ and $k+1$. Thus, the process noise is less in $\omega[k+1]-\omega[k]$ than compared to $\omega[k]$. On other hand, the damping estimator rely more on $\omega[k]$; and hence, its performance is strongly influenced by the process noise. As a result, it requires more measurements to accurately estimate the true damping. 

\begin{figure}
		\centering
		\includegraphics[width=1.0\linewidth]{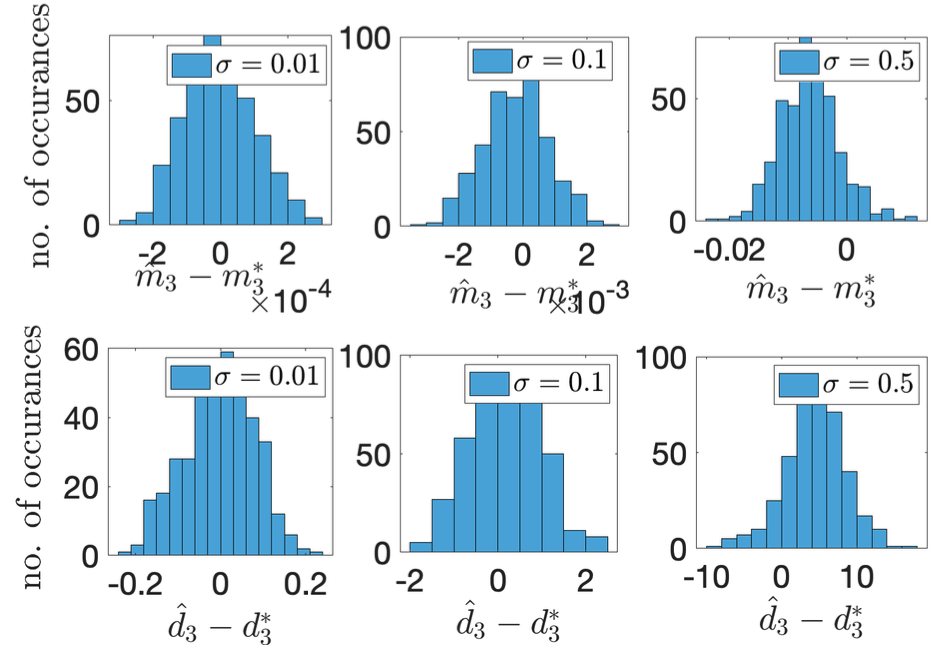} 
		\caption{\small Empirical probability distribution of the error deviation of inertia and damping of generator labelled 3. For $\mathcal{T}=50$, the top panel presents the histograms of error deviations of inertia for various noise levels. The bottom panel presents similar plots for damping. In both the panels, the spread increases (range of x-axis) with increase in $\sigma$. However, this is more pronounced for the case of error deviations of damping constant.} \label{fig: histogram}
	\end{figure}

\begin{figure}
		\centering
		\includegraphics[width=1.0\linewidth]{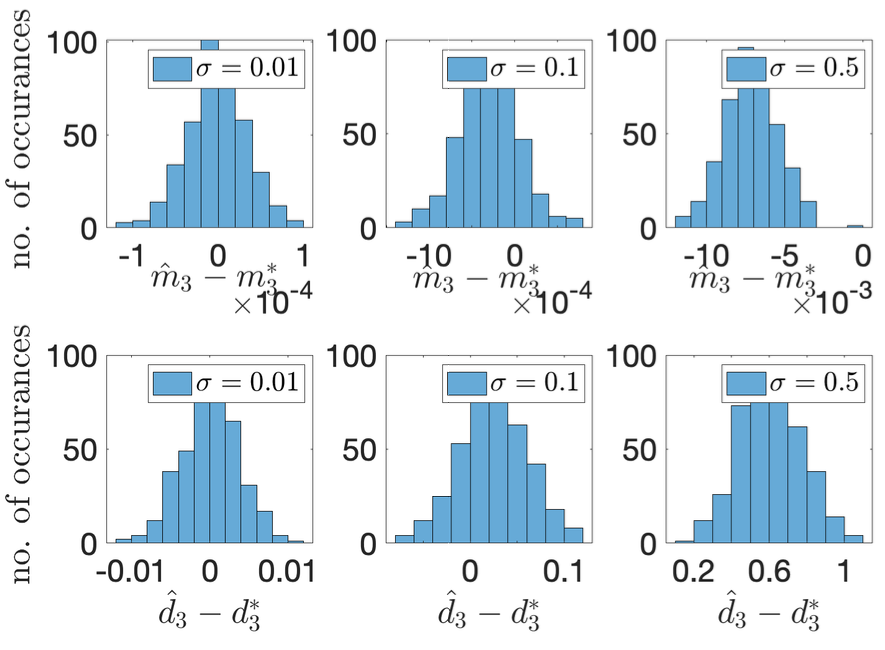} 
		\caption{\small Empirical probability distribution of the error deviation of inertia and damping of generator labelled 3. For $\mathcal{T}=200$, the top panel presents the histograms of error deviations of inertia for various noise levels. The bottom panel presents similar plots for damping. Compared to Fig.~\ref{fig: histogram}, the distribution is more concentrated around zero. This agrees with our intuition that estimation error decreases with the increase in measurements.} \label{fig: histogram2}
	\end{figure}

Our second simulations focus on the probability distribution of the estimation error for a random generator. We chose $i=3$. Fig.~\ref{fig: histogram} and Fig.~\ref{fig: histogram2} illustrate empirical histograms for the estimation time horizons $\mathcal{T}=50$ and $\mathcal{T}=200$. 

\subsection{Case study 2: comparison with the na\"ive esimator in \cite{lokhov2018online}}

Next, we examine the estimator's performance in the presence of both synchronous and converter-faced generators. For the latter, we chose VSMs, whose behavior is emulated by setting the inertia constants to be close (but not exactly) to zero. In particular, we set $m^*_3=0.0019$, $m^*_4=0.0015$, and $m^*_5=0.0014$. We also compared the performance of our estimator with the na\"ive estimator that first estimates $A_d$ in \eqref{eq: DT-state-space1} and then extract the inertia constants. We estimate $A_d$ using the maximum-likelihood technique suggested in \cite{lokhov2018online}. We report our findings in Table \ref{tab:comparsion}. Therein, the values in the parenthesis indicate relative estimation errors. For all the generators, including the VSMs, our structure preserving estimator quite accurately estimated the inertia constants. 

\begin{table}[H]
\normalsize 
    \caption{IEEE 39-bus synchronous generator inertia (p.u.)}
    \vspace{2.0mm}
    \centering
    \begin{tabular}{ |c|c|c|} 
\hline 
True & our method & na\"ive estimator \\
\hline 
$m^*_1$ = 0.2228 & 0.2228 (-0.005e-03) & -0.0384 (-1.1724)\\
$m^*_2$ = 0.1607 & 0.1607 (-0.251e-03) & -0.0014 (-1.0090)\\
$m^*_3$ = {\bf 0.0019} & 0.0019 (-0.042e-03) & -0.0008 (-1.4535)\\
$m^*_4$ = {\bf 0.0015} & 0.0015 (-0.031e-03) & -0.0002 (-0.8677)\\
$m^*_5$ = {\bf 0.0014} & 0.0014 (-0.873e-03) & -0.0002 (-1.1791)\\
$m^*_6$ = 0.1846 & 0.1845 (-0.054e-03) & -0.0915 (-1.4959)\\
$m^*_7$ = 0.1401 & 0.1401 (-0.019e-03) & -0.0864 (-0.3833)\\
$m^*_8$ = 0.1289 & 0.1289 (-0.015e-03) & -0.0144 (-0.8880)\\
$m^*_9$ = 0.1830 & 0.1830 (-0.023e-03) & -0.0369 (-1.2015)\\
$m^*_{10}$ = 2.6526 & 2.6526 (-0.004e-03) & 1.6507 (-0.3777)\\
\hline 
    \end{tabular}
    \label{tab:comparsion}
\end{table}

The simulations presented in this section supported many of our theoretical observations and outperformed methods that do not consider the ill-conditioning aspects as in studies in case 2. 
These observations have implications for design and implementation of real-time algorithms for estimating inertia and damping in low-inertia systems.  

\section{Concluding Remarks}
A simple observation that the parameters of multiple areas or generators could be directly estimated using a descriptor or ill-conditioned electro-mechanical dynamics allowed us to estimate the inertia and damping of power systems with a mix of synchronous and converter-interfaced generators. The latter includes synchronous virtual machines and droop-control-based generators for which the inertia constants are exactly or approximately zero, thereby rendering the utility of the existing inertia and damping estimation methods, which almost always assume non-negligible inertia. We overcome this limitation by studying a constrained least-squares estimator on the descriptor-type dynamics, where the constraints set the inertia of droop-controlled generators to zero. We argued that the proposed estimator is well-posed and admits a unique solution, at least for a special case. Furthermore, we discussed some limitations of the na\"ive estimator in the context of inertia and damping estimation. 

Our analysis highlighted the role of network connectivity on the estimators' performance, which has not been properly studied in the literature. In particular, using the closed-form expressions of the estimators, we showed that for generators with greater connectivity, estimation of the associated parameters is more susceptible to variations in other generator states. Finally, our simulation results showed that estimating the parameters by ignoring the ill-conditioning aspects yields highly unreliable results.



	\bibliographystyle{unsrt}
	\bibliography{BIB.bib}

\end{document}